\documentclass[%
,aps%
 ,twocolumn%
 ,secnumarabic%
,amssymb, amsmath,nobibnotes, aps, prl, floatfix]{revtex4-1}
\usepackage{graphicx}
\usepackage{docs}%
\usepackage{float}
\usepackage{dcolumn}
\usepackage{bm}

\begin{document}

\title{Dipole-Allowed Direct Band Gap Silicon Superlattices}
\author{Young Jun Oh}
\affiliation{Department of Physics, Korea  Advanced Institute of Science and Technology, Daejeon 305-338, Korea}
\author{In-Ho Lee}
\affiliation{Korea Research Institute of Standards and Science, Daejeon 305-340, Korea}
\affiliation{Center for In Silico Protein Science, School of Computational Science, Korea Institute for Advanced Study, Seoul 130-722, Korea}
\author{Sunghyun Kim}
\affiliation{Department of Physics, Korea  Advanced Institute of Science and Technology, Daejeon 305-338, Korea}
\author{Jooyoung Lee}\email[Corresponding author:]{jlee@kias.re.kr}
\affiliation{Center for In Silico Protein Science, School of Computational Science, Korea Institute for Advanced Study, Seoul 130-722, Korea}
\author{K. J. Chang}\email[Corresponding author:]{kjchang@kaist.ac.kr}
\affiliation{Department of Physics, Korea  Advanced Institute of Science and Technology, Daejeon 305-338, Korea}
\date{\today}

\begin{abstract}
Silicon is the most popular material used in electronic devices. However, its poor optical properties owing to its indirect band gap nature limit its usage in optoelectronic devices. Here we present the discovery of super-stable pure-silicon superlattice structures that can serve as promising materials for solar cell applications and can lead to the realization of pure Si-based optoelectronic devices. The structures are almost identical to that of bulk Si except that defective layers are intercalated in the diamond lattice. The superlattices exhibit dipole-allowed direct band gaps as well as indirect band gaps, providing ideal conditions for the investigation of a direct-to-indirect band gap transition. The transition can be understood in terms of a novel conduction band originating from defective layers, an overlap between the valence- and conduction-band edge states at the interface layers, and zone folding with quantum confinement effects on the conduction band of non-defective bulk-like Si. The fact that almost all structural portions of the superlattices originate from bulk Si warrants their stability and good lattice matching with bulk Si. Through first-principles molecular dynamics simulations, we confirmed their thermal stability and propose a possible method to synthesize the defective layer through wafer bonding.
\end{abstract}
 \maketitle

\clearpage

\section{INTRODUCTION}

Silicon is an important element used in modern electronic devices owing to its abundance, feasibility for large-scale fabrication, easy formation of native oxide, and doping controllability of both electrons and holes. However, the optical property of Si is rather poor owing to its indirect electronic band gap nature. In an indirect-band-gap material like cubic-diamond Si (denoted as {\it{c}}-Si), optical transitions at the threshold energy occur only via momentum-conserving phonons. Therefore, the solar spectrum pertaining to the energy below the direct band gap of {\it{c}}-Si, {\it{i.e.}}, approximately 3.4 eV, cannot be effectively absorbed without phonon assistance.

To improve the optical property of Si, considerable efforts have been made; for instance, by introducing defects, such as erbium atoms, dislocations, and grain boundaries as recombination centers \cite{r1,r2,r3}, and/or by engineering the electronic band structure through nanopatterning \cite{r4,r5}, nanostructuring \cite{r6}, alloying with group-IV elements \cite{r7,r8}, applying strain, or combinations of these \cite{r9,r10,r11,r12}. Very recently, several Si crystals with direct and quasidirect band gaps were computationally designed \cite{r13,r14,r15} and their optical absorption properties were shown to be significantly improved compared with {\it{c}}-Si: quasidirect band gaps are defined as $E_g^i \leq E_g^d \leq E_g^i + 0.15$ eV following a previous study \cite{r13}, where $E_g^d$ and $E_g^i$ denote direct and indirect band gaps, respectively. However, these metastable structures were of relatively high energies, ranging 0.1--0.3 eV per atom, because of distorted tetrahedral bonds, and the experimental realization of their synthesis has yet to be explored.

In this work, using a computational search method, we find pure Si-based superlattices that exhibit direct and optically allowed band gaps with good lattice matching with {\it{c}}-Si. The superlattice structure is composed of alternating stacks of bulk-like Si(111) layers and a defective layer containing Seiwatz chains \cite{r16}. The electronic structure evolves as the bulk-like Si portion increases, exhibiting a transition from the direct to indirect band gap. In superlattices with direct band gaps, the optical transition at the threshold energy is greatly enhanced. Hence we suggest that these superlattices could be used in thin-film solar cells. In the next section, we explain the details of the computational method. Then we present the structural, electronic, and optical properties of our designed superlattices. Finally, we discuss the thermal stability and the possible synthesis through wafer bonding, based on molecular dynamics simulations.

\begin{figure*}
\centering
\includegraphics[width=6.8in]{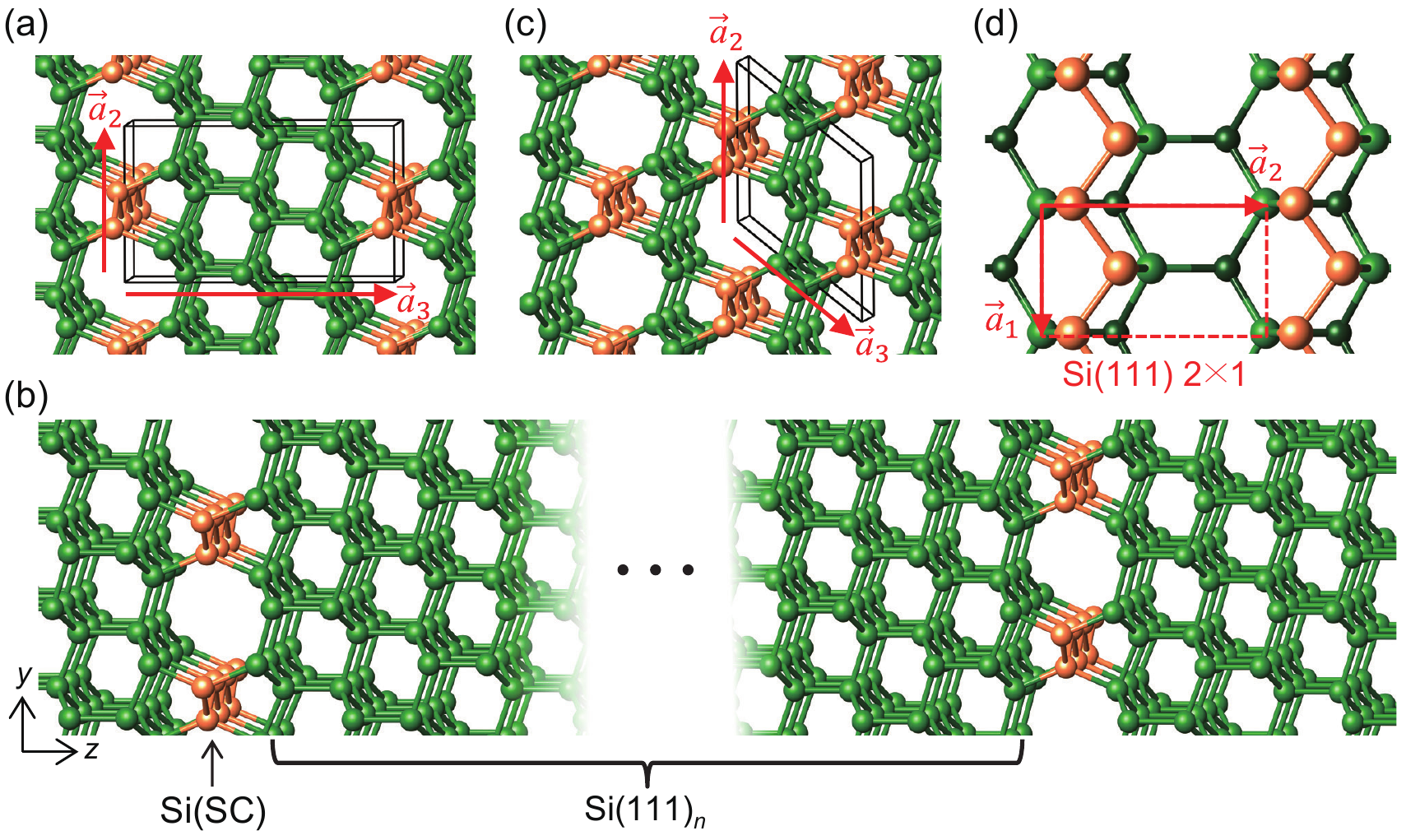}
\caption{\label{fig1} Superlattices composed of Si(111) layers (green) and a defective layer of Seiwatz chains (orange) are shown for (a) $n$ = 3, (b) arbitrary $n$, and (c) $n$ = 1. The black parallelepiped represents the unit cell spanned by the lattice vectors, $\vec{a}_1$, $\vec{a}_2$, and $\vec{a}_3$. (d) The orientation of Seiwatz chains on the Si(111) 2$\times$1 surface is shown.
}
\end{figure*}

\section{CALCULATION METHOD}

We explored Si crystal structures with optically active direct band gaps by using the computational search method \cite{r17}, in which electronic properties are initially assigned and target materials are subsequently searched. We employed a combined approach \cite{r13} of conformational space annealing (CSA) \cite{r18,r19,r20,r21} for global optimization and first-principles calculations within the density functional theory (DFT) framework. Without using any specific knowledge of known Si crystals, we optimized the degrees of freedom including atomic positions $\left\lbrace \vec{R}_I\right\rbrace$ and six lattice parameters ($a$, $b$, $c$, $\alpha$, $\beta$, and $\gamma$). The objective function used for selection in CSA is  $-E_g^i + \textrm{max} \left[ 0, (E_g^d-E_g^i) \right]$, which promotes the formation of direct band gap. For each conformation, the enthalpy was minimized by performing DFT calculations which used the functional form of Perdew, Burke, and Ernzerhof (PBE) \cite{r22} for the exchange-correlation potential and the projector augmented wave potentials \cite{r23}, as implemented in the VASP code \cite{r24}. The wave functions were expanded in plane waves with an energy cutoff of 400 eV. With a $\vec{k}$-point mesh using the grid spacing of $2\pi\times0.02$ {\AA}$^{-1}$, the crystal structures were minimized until all forces and stress tensors were less than 0.01 eV/{\AA} and 1.5 kbar, respectively. Finally, a twice finer $\vec{k}$-point mesh was used to determine the nature of band gap (direct vs indirect). 

The band gap sizes of semiconductors and insulators are usually underestimated with the PBE functional. For more accurate calculations for the band gaps and absorption coefficients, we additionally performed quasiparticle calculations in the {\it{G}}$_0${\it{W}}$_0$ approximation \cite{r25,r26} and solved the Bethe-Salpeter equation \cite{r27}, considering up to 12 occupied and 16 unoccupied bands around the Fermi level, which were shown to be sufficient to ensure the numerical convergence \cite{r13}.

\section{RESULTS AND DISCUSSION}

\begin{table*}
\caption{For the Si(111)$_n$/Si(SC) superlattices with the cubic- and hexagonal-stacking sequences of the Si(111) layers, the lattice type, the number of atoms per unit cell ($N$), the energy relative to cubic-diamond Si ($E$), the type of band gap, the direct band gap size ($E_g^d$), and the indirect band gap size ($E_g^i$) are shown, based on the PBE calculations. The quasiparticle {\it{G}}$_0${\it{W}}$_0$ gaps are also shown for comparison. Here D and QD in parentheses denote direct and quasidirect band gaps, respectively, and lattice types are abbreviated, such as SM: simple monoclinic, SO: simple orthorhombic, and BCO: base-centered orthorhombic.
}
\label{table1}
\begin{ruledtabular}
\begin{tabular}{cccccccc}
$n$ & lattice & $N$ & $E$ (meV/atom) & PBE-$E_g^d$ (eV) & PBE-$E_g^i$ (eV) & {\it{G}}$_0${\it{W}}$_0$-$E_g^d$ (eV) & {\it{G}}$_0${\it{W}}$_0$-$E_g^i$ (eV) \\ \hline
1  & BCO & 6 & 89 (QD) & 0.431 & 0.430 & 0.894 & 0.847 \\ 
\multicolumn{8}{c}{Cubic-diamond stacking} \\
2  & SM & 10 & 46 (QD) & 0.906 & 0.869 & 1.346 & 1.316 \\ 
3  & SM & 14 & 42 (D) & 0.807 &       & 1.197 &       \\ 
4  & SM & 18 & 32 (D) & 0.832 &       & 1.283 &       \\ 
5  & SM & 22 & 26 (D) & 0.782 &       & 1.218 &       \\ 
6  & SM & 26 & 22 (QD) & 0.788 & 0.774 & 1.224 & 1.215 \\ 
7  & SM & 30 & 19 (QD) & 0.761 & 0.741 & 1.198 & 1.185 \\ 
8  & SM & 34 & 17 (QD) & 0.746 & 0.726 & 1.185 & 1.166 \\ 
9  & SM & 38 & 16 (QD) & 0.738 & 0.712 & 1.177 & 1.154 \\ 
10 & SM & 42 & 13 (QD) & 0.725 & 0.699 & 1.167 & 1.147 \\ 
\multicolumn{8}{c}{Hexagonal-diamond stacking} \\
2  & SO & 10 & 72 (QD) & 0.562 & 0.527 & 0.980 & 0.961 \\ 
3  & SM & 14 & 49 (D) & 0.615 &       & 1.049 &       \\ 
4  & SO & 18 & 30 (D) & 0.578 &       & 1.008 &       \\ 
5  & SM & 22 & 34 (D) & 0.497 &       & 0.914 &       \\ 
\end{tabular}
\end{ruledtabular}
\end{table*}

After an extensive search for Si crystal structures with optically active direct band gaps, we obtained a very distinctive superlattice structure, especially for the system containing 14 Si atoms per unit cell. The structure consists of alternating stacks of three Si(111) layers and a defective layer along the [111] direction of {\it{c}}-Si [Fig.~\ref{fig1}(a)]. The defective layer contains the so-called Seiwatz chains (SCs) \cite{r16}, which are known to be formed via 2$\times$1 reconstruction on the Si(111) surface with half-monolayer coverage. The SCs lead to open channels consisting of five- and eight-membered rings in the defective region. Note that all Si atoms are four-fold coordinated, without any coordination defects, and nearly ideal tetrahedral bonds are formed in the non-defective region. Owing to the almost identical structural match between the Si(111) layers and {\it{c}}-Si, many additional superlattice structures [denoted as Si(111)$_n$/Si(SC)] can be constructed by varying $n$, as shown in Fig.~\ref{fig1}(b). For $n = 1$, the crystal is entirely composed of five- and eight-membered rings [Fig.~\ref{fig1}(c)], very similar to an orthorhombic allotrope of Si, Si$_{24}$, which has been recently synthesized \cite{r28}.

In the cubic-diamond (Si-I) phase, the stacking sequence of the Si(111) layers is $ABCABC...$, while that of the hexagonal-diamond (lonsdaleite, Si-IV) phase is $ABABAB...$. In our superlattice system, various stacking sequences including the cubic and hexagonal ones are possible, as in SiC polytypes. In addition, in each defective layer, two configurations of SCs exist, which are related to each other by the translation of $(\vec{a}_1 + \vec{a}_2 )/2$, where $\vec{a}_1$ and $\vec{a}_1$ are the lattice vectors on the 2$\times$1 basal plane [Fig.~\ref{fig1}(d)]. Thus, depending on the relative positions of two adjacent defective layers, two Bravais lattices can be formed, simple monoclinic (SM) and base-centered monoclinic (BCM) for $n \geq 2$, with some exceptions for the case of hexagonal stacking (Table~\ref{table1}). By examining various superlattices, we found that BCM superlattices tend to yield quasidirect/indirect band gaps, whereas the nature of the band gaps (direct vs quasidirect) in the SM ones varies with $n$.

\begin{figure}
\centering
\includegraphics[width=3.2in]{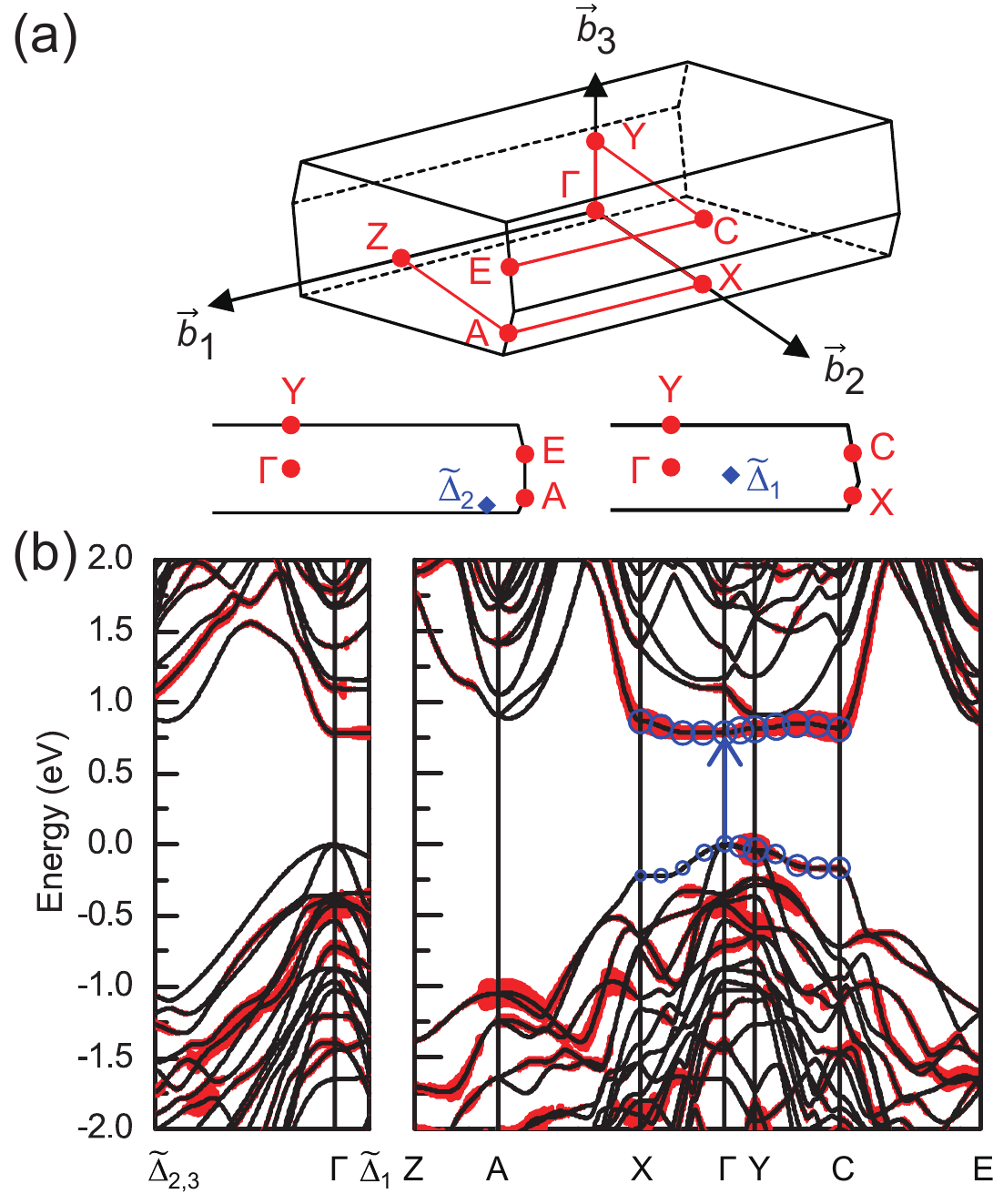}
\caption{\label{fig2} (a) Symmetry points and lines in the Brillouin zone of the simple monoclinic lattice are shown. (b) The PBE band structure of the Si(111)$_{n=5}$/Si(SC) superlattice is shown. The thickness of red colored bands represents the degree of confinement in the defective layer, indicating that the lowest conduction band along the X-$\Gamma$-Y-C line is mainly derived from the Seiwatz chains. For the highest valence and lowest conduction bands along the X-$\Gamma$-Y-C line, the size of blue circles is proportional to the degree of contribution from the interface layer.
}
\end{figure}

\begin{figure}
\centering
\includegraphics[width=3.2in]{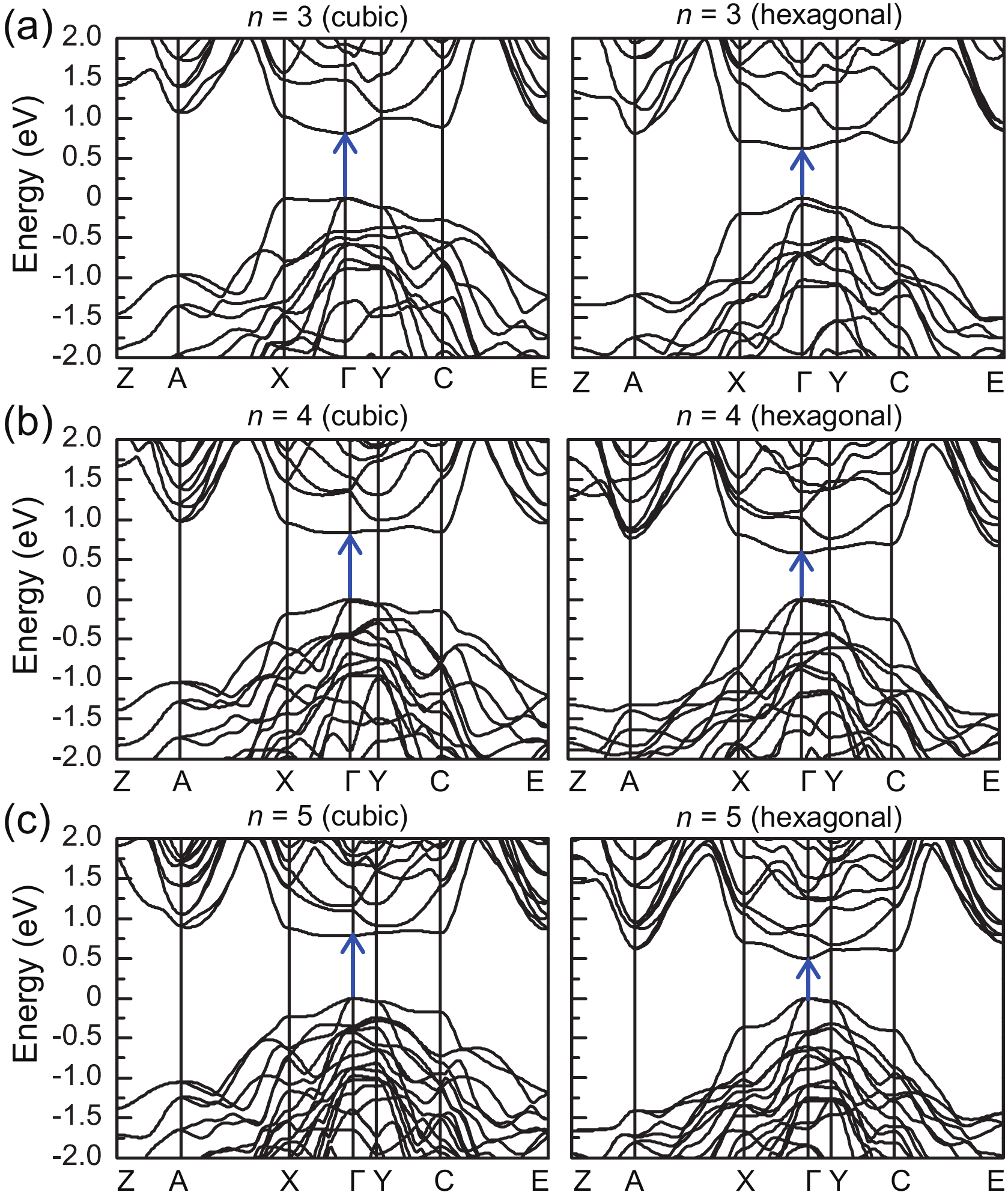}
\caption{\label{fig3} (a)-(c) The PBE band structures are compared for the Si(111)$_n$/Si(SC) superlattices ($n$ = 3--5) with the cubic- and hexagonal-stacking sequences of the Si(111) layers.
}
\end{figure}

Because {\it{c}}-Si is the most stable form of Si, we focused on SM superlattices with cubic-diamond stacking; throughout this paper, unless otherwise specified, the cubic stacking is considered. For superlattices with $n$ up to $n = 13$, the valence band maximum (VBM) is always located at the $\Gamma$ point, the center of the Brillouin zone (BZ). For $n$ = 3--5, we found direct band gaps at $\Gamma$ (Figs.~\ref{fig2} and \ref{fig3}), which were estimated to be 0.807, 0.832, and 0.782 eV, respectively, using the PBE functional for the exchange-correlation potential (Table~\ref{table1}). As $n$ increased, a direct-to-quasidirect band gap transition occurred. The energy differences between the direct and indirect band gaps were smaller than 30 meV for $6 \leq n \leq 10$. Although the band gap size decreased with increasing $n$, it was larger than that (0.62 eV) of {\it{c}}-Si, owing to the quantum confinement effect. When more rigorous quasiparticle {\it{G}}$_0${\it{W}}$_0$ calculations were performed, the nature of the band gap did not change. The {\it{G}}$_0${\it{W}}$_0$ band gaps were 1.15--1.28 eV ($n$ = 3--10), close to the optimal value (1.1 or 1.3 eV) for solar cell applications \cite{r29} (Table~\ref{table1}). For comparison, we also examined SM superlattices with hexagonal-diamond stacking and found direct band gaps at $\Gamma$ for $n$ = 3--5. However, the band gap sizes were reduced by approximately 0.19--0.29 eV (Table~\ref{table1}), similar to SiC polytypes, where the band gap tends to decrease as the number of hexagonal layers in stacking sequence increases.

\begin{figure}
\centering
\includegraphics[width=3.2in]{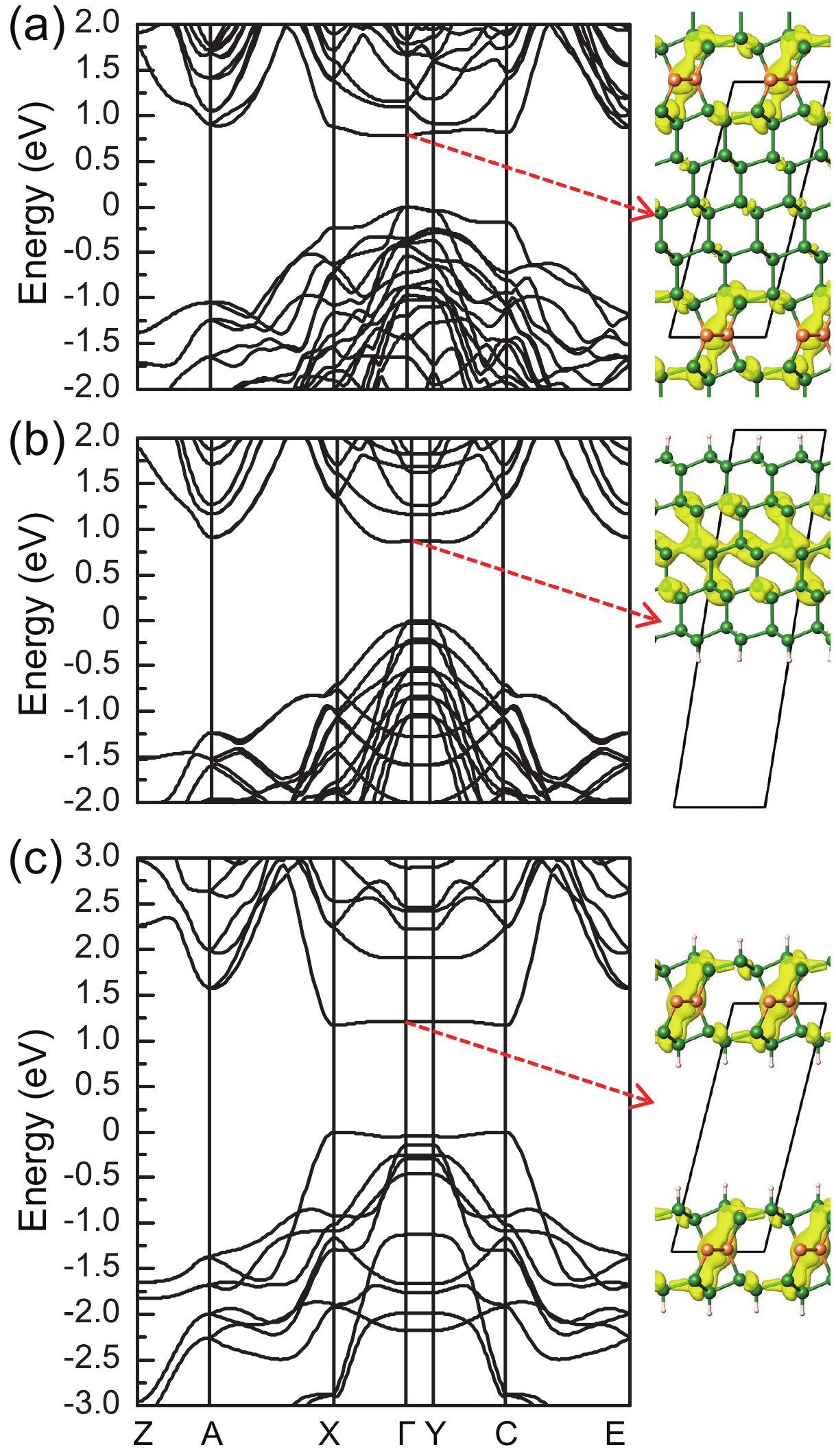}
\caption{\label{fig4} The PBE band (left panel) and atomic (right panel) structures are compared for (a) the Si(111)$_{n=5}$/Si(SC) superlattice with the cubic-stacking sequence of the Si(111) layers, (b) a slab geometry consisting of five Si(111) layers and a vacuum region, and (c) a slab geometry consisting of a defective layer sandwiched between two Si(111) layers and a vacuum region. Surface Si dangling bonds are passivated by hydrogen. In right panel, isosurfaces (yellow) of the charge densities of the lowest conduction bands at the $\Gamma$ point are drawn.
}
\end{figure}

\begin{figure}
\centering
\includegraphics[width=3.2in]{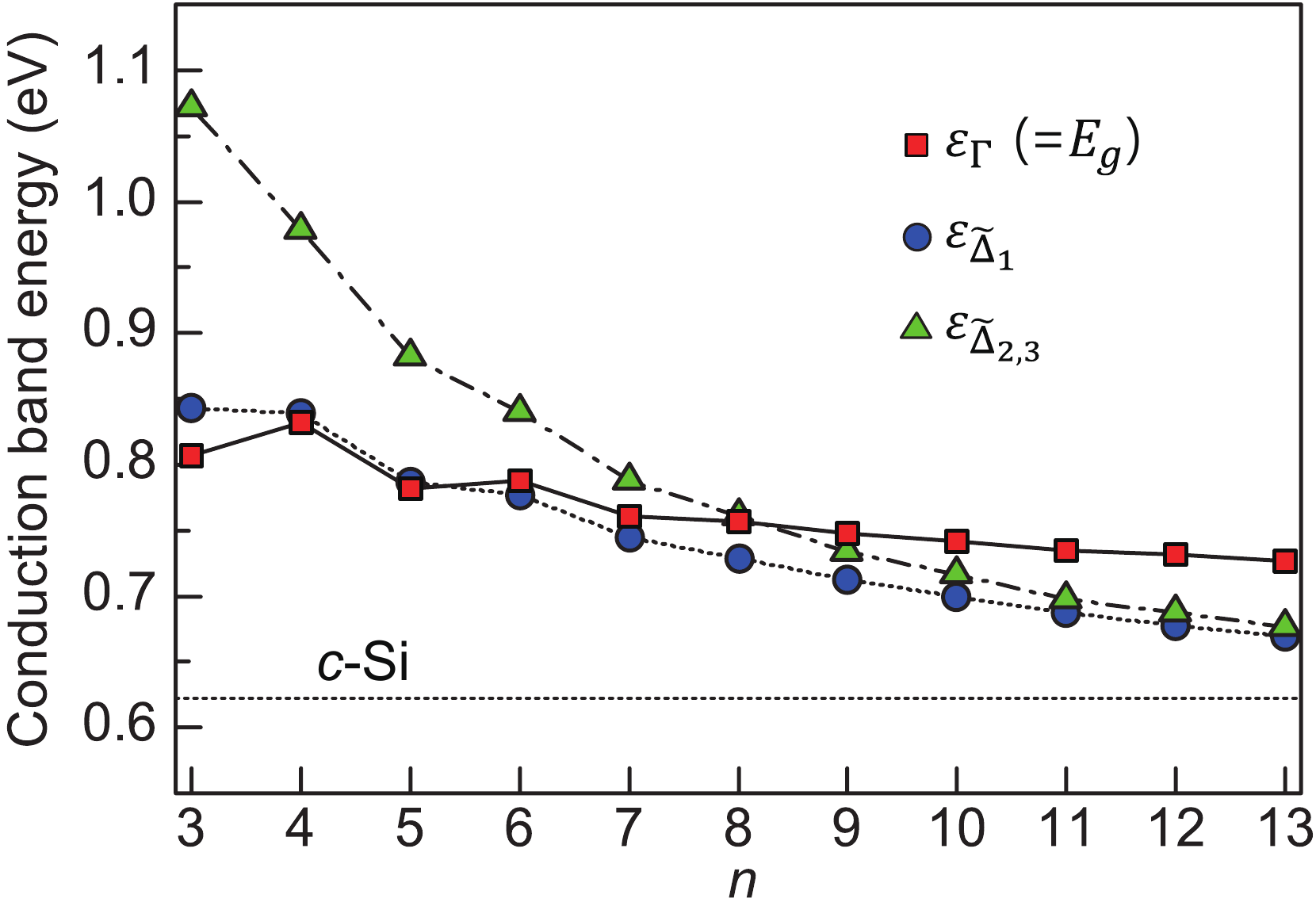}
\caption{\label{fig5} The lowest conduction band energies at $\Gamma$, $\tilde{\Delta}_1$, and $\tilde{\Delta}_{2,3}$ are plotted as a function of $n$ and compared with that of {\it{c}}-Si (horizontal dotted line).
}
\end{figure}

In our superlattice system, the lowest conduction band exhibits nearly flat dispersion along the $\Gamma$-X and $\Gamma$-Y directions (Figs.~\ref{fig2} and \ref{fig3}). To understand the origin of the flat dispersion around $\Gamma$, we considered the defective and Si(111) layers separately (Fig.~\ref{fig4}). In a slab geometry with a single defective layer sandwiched between two Si(111) layers, a vacuum region was included in the same unit cell and Si dangling bonds were passivated by hydrogen. It is clear that the flat dispersion is attributed to the SCs in the defective layer, as shown in Fig.~\ref{fig4}. On the other hand, the energy states derived from the Si(111) layers are affected by the zone folding and quantum confinement effects. In {\it{c}}-Si, the conduction band minimum (CBM) is located at six $\Delta$-valleys close to the X-points in the BZ of the face-centered cubic lattice. When a SM cell with a 2$\times$1 basal plane is adopted, two of the six X-points are folded to $\Gamma$ in the monoclinic BZ. In contrast to {\it{c}}-Si, the hexagonal symmetry in the superlattice is perturbed by the SCs [Fig.~\ref{fig1}(d)]; hence, six $\Delta$-valley states split into two folded states, $\tilde{\Delta}_1$ and  $\tilde{\Delta}_{2,3}$, which have two- and four-fold degeneracies, respectively. The folded $\tilde{\Delta}_1$ points are located near $\Gamma$, whereas the folded $\tilde{\Delta}_{2,3}$ points are close to A or E in BZ, depending on $n$ [Fig.~\ref{fig2}(a)]. The lowest conduction bands at $\Gamma$, $\tilde{\Delta}_1$, and $\tilde{\Delta}_{2,3}$ are shown as a function of $n$ in Fig.~\ref{fig5}. While the $\tilde{\Delta}_{2,3}$ states are mainly confined in the Si(111) layers, reflecting their zone folding nature, the characteristics of $\tilde{\Delta}_1$ vary with $n$.

\begin{figure}
\centering
\includegraphics[width=3.0in]{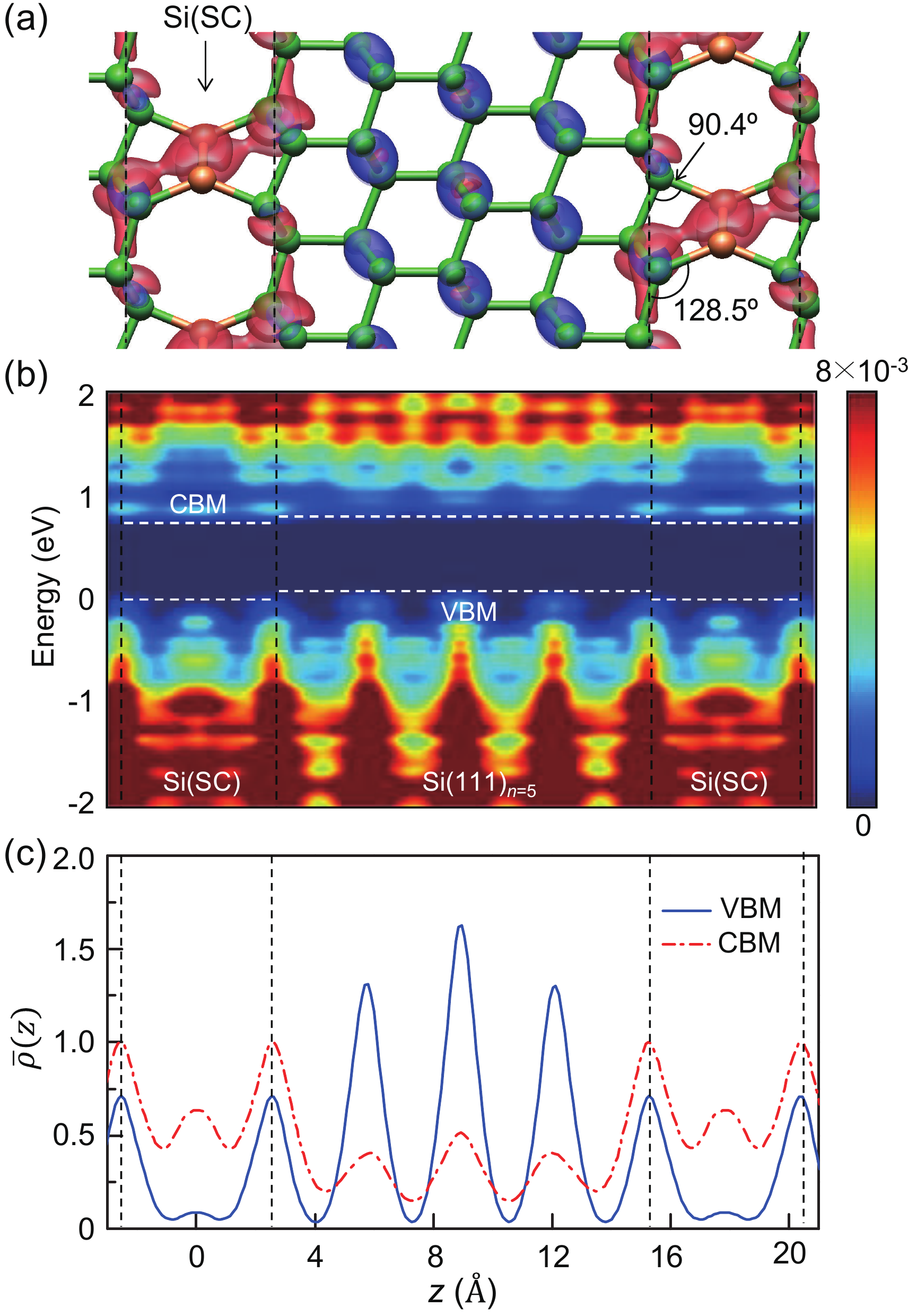}
\caption{\label{fig6} For the Si(111)$_{n=5}$/Si(SC) superlattice, (a) isosurfaces ($1.32\times10^{‒2}$ electrons/{\AA}$^3$) of the charge densities of VBM (blue) and CBM (red), (b) local density of states (in units of electrons/eV/{\AA}$^3$) averaged over the xy plane, and (c) planar-averaged charge densities of VBM and CBM are plotted along the superlattice direction ($z$-axis). Black dotted lines denote the position of interface Si(111) layers and white dotted lines represent the approximate positions of the band edge states in the middle of the defective and non-defective regions.
}
\end{figure}

The nature of the band gap in our superlattices was determined by the competition between the novel defect-derived flat state and the folded states of the Si(111) layers in the conduction band. For small $n$, the band gap of the non-defective region becomes large due to the quantum confinement effect. Thus, the defect-derived flat band at $\Gamma$ is lower than all the folded states, exhibiting a direct band gap behavior up to $n$ = 5. The planar-averaged charge densities clearly show that CBM is largely confined in the defective region, while VBM is mainly derived from the Si(111) layers (Fig.~\ref{fig6}). For large $n$, as the confinement effect is reduced, the folded $\tilde{\Delta}_1$ states move down below the defect-derived flat band, resulting in a direct-to-quasidirect band gap transition around $n$ = 6 (Fig.~\ref{fig5}).

\begin{figure}
\centering
\includegraphics[width=3.2in]{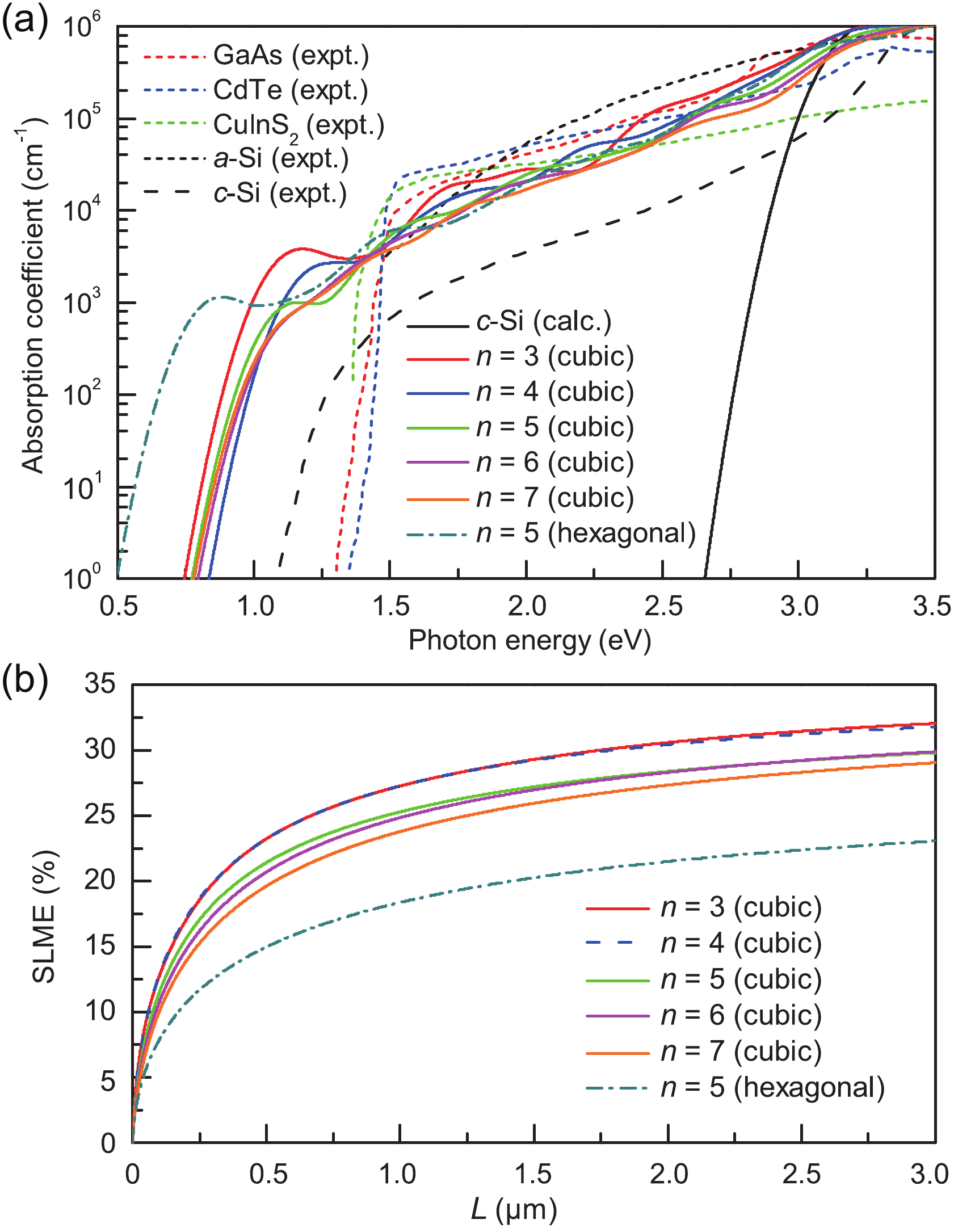}
\caption{\label{fig7} For various Si(111)$_n$/Si(SC) superlattices with the cubic- and hexagonal-stacking sequences of the Si(111) layers, (a) the calculated absorption coefficients are compared with the experimentally measured values (from Refs. \cite{r30,r31,r32,r33}) for GaAs, CdTe, CuInS$_2$, amorphous Si ({\it{a}}-Si), and {\it{c}}-Si, and (b) the spectroscopic limited maximum efficiency (SLME) is plotted as a function of film thickness $L$.
}
\end{figure}

The novel defect-derived band plays an important role in strong optical transitions near the threshold energy. The squares of the dipole matrix elements, $\lvert\left\langle f | \vec{p}|i \right\rangle \rvert^2$, for the direct transition at $\Gamma$ were calculated to be 0.173, 0.141, and 0.045 $a_0^{-2}$ in atomic units for $n$ = 3--5, respectively, where $a_0$ is the Bohr radius. These values are higher than 0.03 $a_0^{-2}$ obtained from the dipole-allowed direct band gap of a specially designed Si/Ge superstructure [11], indicating that the optical transition was greatly enhanced in our case. Such strong dipole-allowed transitions are partly attributed to the large overlap of the band edge states at the interface layers (Figs.~\ref{fig2} and \ref{fig6}). In addition, the hybridization of the conduction band states is another critical factor for the dipole-allowed transition. In {\it{c}}-Si, the band edge states at $\Gamma$ are characterized by bonding and antibonding $p$ orbitals and the optical transition at the direct gap is dipole-allowed. In our superlattices with direct band gaps, we examined the orbital characteristics by projecting the wave function onto the atom-centered spherical harmonics within a sphere of 1.5 {\AA} radius. In flanking Si(111) layers, tetrahedral bonds, especially around the interface Si atoms bonded to the SCs, are distorted from their ideal values, with the bond angles of 90.4$^\circ$ and 128.5$^\circ$ (Fig.~\ref{fig6}). As a consequence, the $p$ orbital character is significantly enhanced for the CBM state, as in the dipole-allowed transition at the direct gap of {\it{c}}-Si. The calculated absorption coefficients of our superlattices are comparable to those of direct band gap semiconductors, such as GaAs, CdTe, and CuInS$_2$ [Fig.~\ref{fig7}(a)], which are known as good photovoltaic materials. The spectroscopic limited maximum efficiency \cite{r34} for the sample thickness of $L$ = 2.0 $\mu$m was estimated to be in the range of 27\%--31\% for $n$ = 3--7 [Fig.~\ref{fig7}(b)], indicating that the optical absorption properties were excellent even for the quasidirect band gap superlattices.

\begin{figure}
\centering
\includegraphics[width=3.2in]{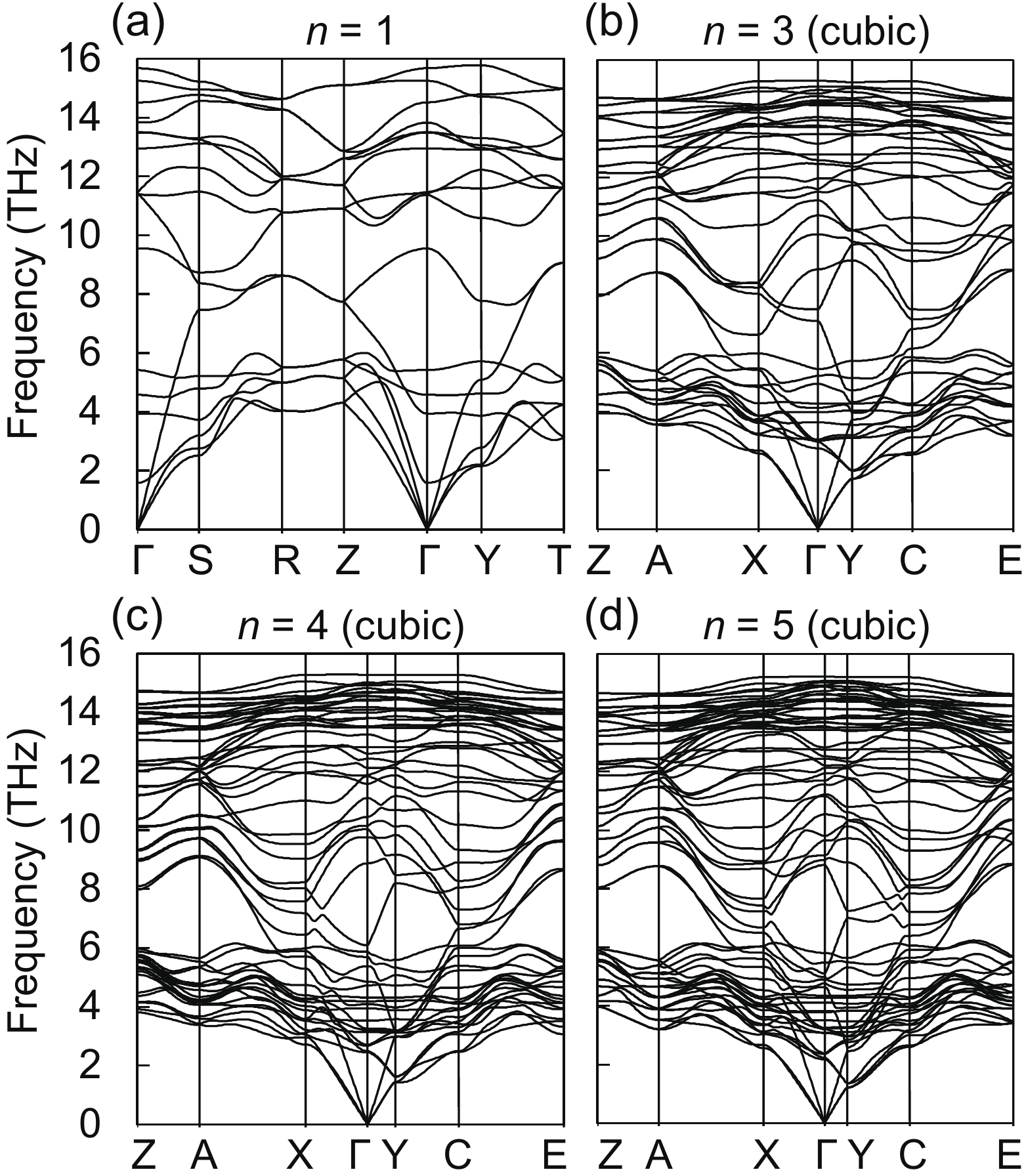}
\caption{\label{fig8} For the Si(111)$_n$/Si(SC) superlattices with the cubic-stacking sequence of the Si(111) layers, the calculated phonon spectra are shown for $n$ = 1, 3, 4, and 5. The $2\times2\times3$, $3\times2\times2$, $3\times2\times2$, and $3\times2\times2$ supercells are used for $n$ = 1, 3, 4, and 5, respectively.
}
\end{figure}

\begin{figure}
\centering
\includegraphics[width=3.2in]{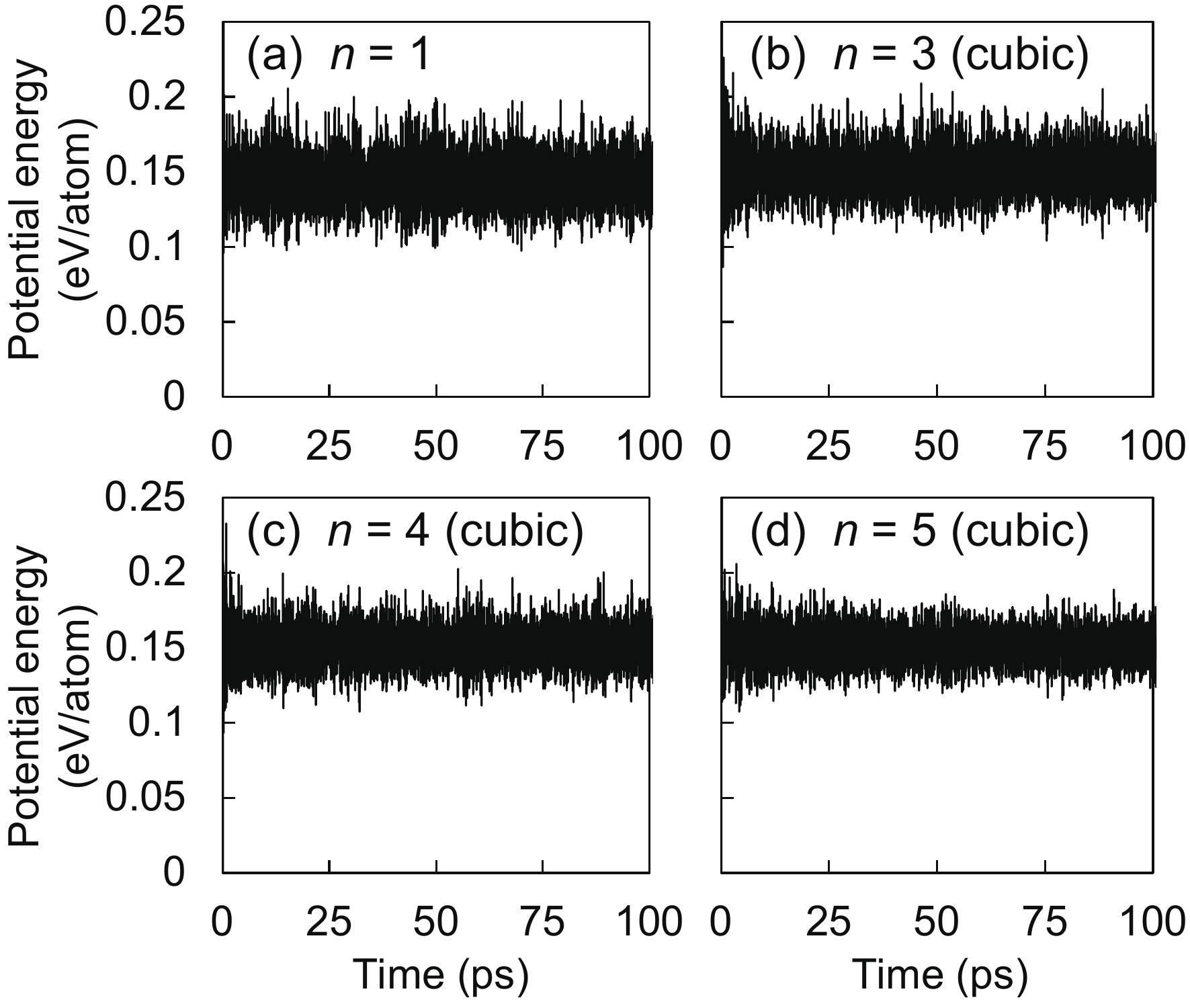}
\caption{\label{fig9} For the Si(111)$_n$/Si(SC) superlattices with the cubic-stacking sequence of the Si(111) layers, the thermal stability was examined by performing first-principles MD simulations at 1100 K, with choosing the $2\times2\times3$, $3\times2\times1$, $3\times2\times1$, and $3\times2\times1$ supercells for $n$ = 1, 3, 4, and 5, respectively.
}
\end{figure}

Note that our superlattices are both energetically and dynamically stable at the ambient condition. Since the Si(111) layers are structurally almost identical to {\it{c}}-Si, their excess energies are quite low, ranging 0.013--0.042 eV per atom for $n$ = 3--10 (Table~\ref{table1}). These energies are lower by an order of magnitude than those (0.1--0.3 eV per atom) of the previously predicted Si allotropes with direct and quasidirect band gaps \cite{r13,r14,r15}. For our direct band gap superlattices, we found no imaginary phonon modes in the phonon spectra (Fig.~\ref{fig8}). In addition, we examined their thermal stability by performing first-principles molecular dynamics (MD) simulations and confirmed that they were thermally stable up to 100 ps at the high temperature of 1100 K (Fig.~\ref{fig9}).

\begin{figure}
\centering
\includegraphics[width=3.2in]{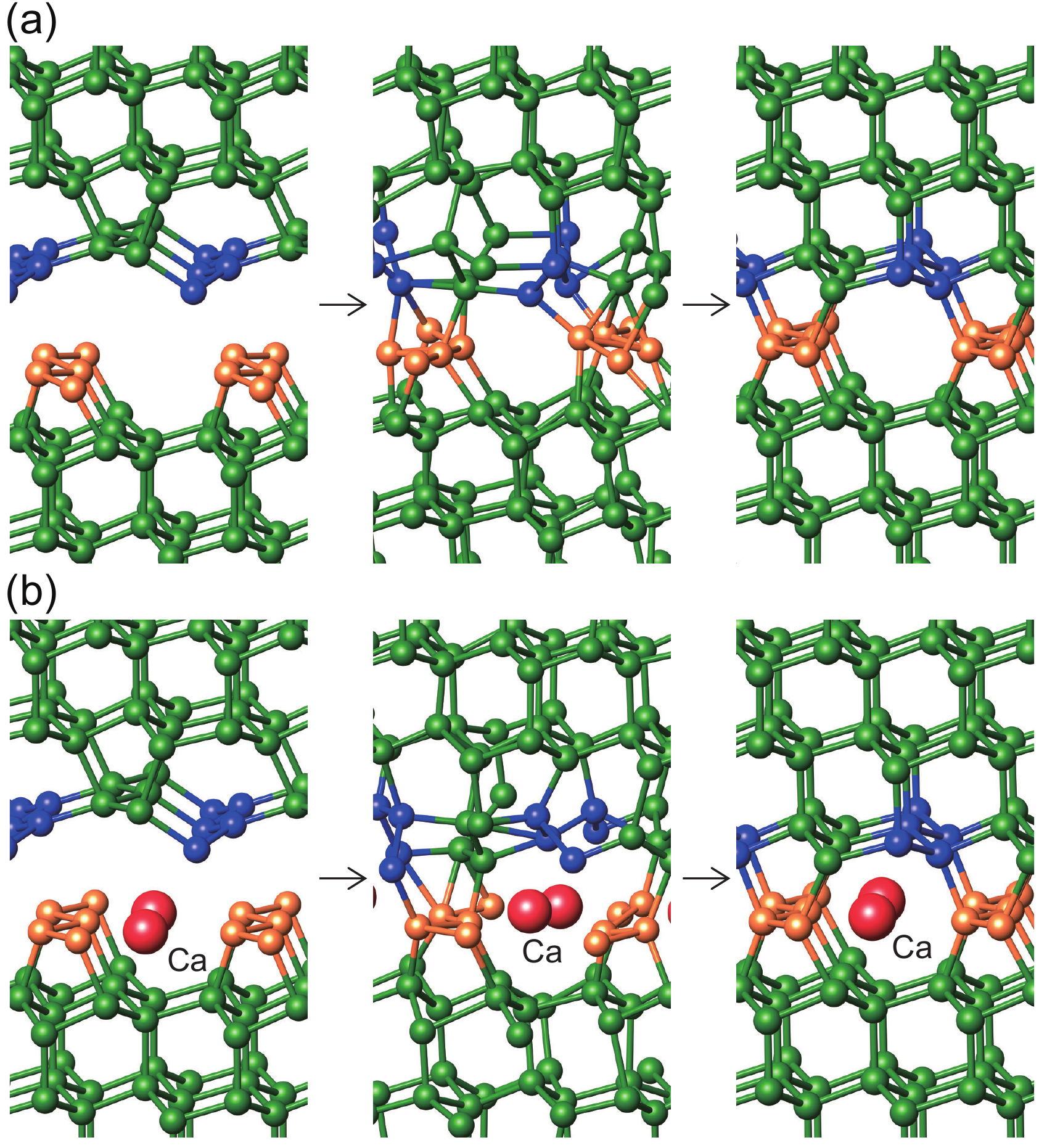}
\caption{\label{fig10} (a) For the initial configuration where one surface has Pandey $\pi$-bonded chains (blue circles) and the other has Seiwatz chains (orange circles), an intermediate configuration after 2 ps and the final configuration after 3 ps at 1100 K are shown during molecular dynamics simulations for wafer bonding. (b) For the initial configuration where Ca atoms are adsorbed between Seiwatz chains on one surface, an intermediate configuration after 3 ps and the final configuration after 7 ps at 1100 K are shown. In the final state, the Ca atoms are aligned along the open channels of eight-membered rings.
}
\end{figure}

Finally, we examined the possibility of creating an SC layer, which can lead to the eventual realization of our superlattice structures. We performed first-principles MD simulations for wafer bonding between two Si(111) 2$\times$1 surfaces. For the initial configuration, where one surface had Pandey $\pi$-bonded chains \cite{r35} and the other SCs \cite{r16}, a 4$\times$2 lateral supercell was chosen, with a vacuum region of 15.0 {\AA} inserted between the two surfaces passivated by hydrogen. It was found that a defective layer with the SCs was formed after about 3 ps at 1100 K [Fig.~\ref{fig10}(a)]. Here, the 2$\times$1 reconstruction of the SCs may not be possible, since it is metastable with respect to the Pandey reconstruction model. On the other hand, divalent adsorbates, such as Ca, Sr, and Ba, can stabilize the SCs at half-monolayer coverage \cite{r36}. Thus, we considered one of the 2$\times$1 surfaces with Ca atoms adsorbed at hollow surface sites [Fig.~\ref{fig10}(b)] and found that the same defective layer was formed after approximately 7 ps. In the final configuration, the Ca adsorbates resided along the open channels of eight-membered rings, as in the case of the Eu$_4$Ga$_8$Ge$_{16}$-type structure found in CaSi$_6$, SrSi$_6$, BaSi$_6$, \cite{r37} and recently in NaSi$_6$ \cite{r28,r38}. Once the defective layer is formed, the adsorbed Ca atoms can be removed via a diffusion process along the channel, as experimentally observed in the case of NaSi$_6$ \cite{r28}. Thus, the wafer bonding between the clean and divalent-adsorbed Si(111) surfaces can serve as a promising technique for the synthesis of superlattices containing the defective SC layer.

\section{CONCLUSIONS}
Using a computational search method, we have predicted low-energy pure-Si superlattice structures with dipole-allowed direct band gaps, which can serve as promising materials for solar-cell applications. In the superlattice structure, the defective layers containing Seiwatz chains are intercalated in the diamond lattice of bulk Si. A direct-to-indirect band gap transition occurs as the bulk-like Si portion increases, which is understood in terms of a defect-originated conduction band and zone folding with quantum confinement effects on the competing conduction band of bulk-like Si layers. The method proposed for the synthesis of the defective layer through wafer bonding can be followed up by experiments and could herald a new optoelectronic era.

\begin{acknowledgments}

\end{acknowledgments}
IHL and JL are supported by the National Research Foundation of Korea (NRF) under Grant No. 2008-0061987 funded by the Korea government (MEST). YJO, SK, IHL, and KJC are supported by Samsung Science and Technology Foundation under Grant No. SSTF-BA1401-08. We thank Korea Institute for Advanced Study (KIAS Center for Advanced Computation) for providing computing resources.


\begin{references}

\bibitem{r1}
B. Zheng, J. Michel, F. Y. G. Ren, L. C. Kimerling, D. C. Jacobson, and J. M. Poate, {\it{Room-Temperature Sharp Line Electroluminescence at $\lambda=1.54$ $\mu$m from an Erbium-Doped, Silicon Light-Emitting Diode}}, Appl. Phys. Lett. {\textbf{64}}, 2842 (1994).

\bibitem{r2}
W. L. Ng, M. A. Louren\c{c}o, R. M. Gwilliam, S. Ledain, G. Shao, and K. P. Homewood, {\it{An Efficient Room-Temperature Silicon-Based Light-Emitting Diode}}, Nature {\textbf{410}}, 192 (2001).

\bibitem{r3}
R. Raghunathan, E. Johlin, and J. C. Grossman, {\it{Grain Boundary Engineering for Improved Thin Silicon Photovoltaics}}, Nano Lett. {\textbf{14}}, 4943 (2014).

\bibitem{r4}
S. G. Cloutier, P. A. Kossyrev, and J. Xu, {\it{Optical Gain and Stimulated Emission in Periodic Nanopatterned Crystalline Silicon}}, Nature Mater. {\textbf{4}}, 887 (2005).

\bibitem{r5}
D. Li, L. Lin, and J. Feng, {\it{Electronic State and Momentum Matrix of H-Passivated Silicon Nanonets: A first-principles calculation}}, Physica E {\textbf{42}}, 1583 (2010).

\bibitem{r6}
Y. Shirasaki, G. J. Supran, M. G. Bawendi, and V. Bulovi\c{c}, {\it{Emergence of Colloidal Quantum-Dot Light-Emitting Technologies}}, Nature Photon. {\textbf{7}}, 13 (2012).

\bibitem{r7}
J. Weber and M. I. Alonso, {\it{Near-Band-Gap Photoluminescence of Si-Ge Alloys}}, Phys. Rev. B {\textbf{40}}, 5683 (1989).

\bibitem{r8}
P. Zhang, V. H. Crespi, E. Chang, S. G. Louie, and M. L. Cohen, {\it{Computational Design of Direct-Bandgap Semiconductors That Lattice-Match Silicon}}, Nature {\textbf{409}}, 69 (2001).

\bibitem{r9}
B. Huang, H.-X. Deng, H. Lee, M. Yoon, B. G. Sumpter, F. Liu, S. C. Smith, and S.-H. Wei, {\it{Exceptional Optoelectronic Properties of Hydrogenated Bilayer Silicene}}, Phys. Rev. X {\textbf{4}}, 021029 (2014).

\bibitem{r10}
K. K\r{u}sov\'{a}, P. Hapala, J. Valenta, P. Jel\'{i}nek, O. Cibulka, L. Ondi\v{c}, and I. Pelant, {\it{Direct Bandgap Silicon: Tensile-Strained Silicon Nanocrystals}}, Adv. Mater. Interfaces {\textbf{1}}, 1300042 (2014).

\bibitem{r11}
M. d'Avezac, J.-W. Luo, T. Chanier, and A. Zunger, {\it{Genetic-Algorithm Discovery of a Direct-Gap and Optically Allowed Superstructure from Indirect-Gap Si and Ge Semiconductors}}, Phys. Rev. Lett. {\textbf{108}}, 027401 (2012).

\bibitem{r12}
L. Zhang, M. d’Avezac, J.-W. Luo, and A. Zunger, {\it{Genomic Design of Strong Direct-Gap Optical Transition in Si/Ge Core/Multishell Nanowires}}, Nano Lett. {\textbf{12}}, 984 (2012).

\bibitem{r13}
I.-H. Lee, J. Lee, Y. J. Oh, S. Kim, and K. J. Chang, {\it{Computational Search for Direct Band Gap Silicon Crystals}}, Phys. Rev. B {\textbf{90}}, 115209 (2014).

\bibitem{r14}
Q. Wang, B. Xu, J. Sun, H. Liu, Z. Zhao, D. Yu, C. Fan, and J. He, {\it{Direct Band Gap Silicon Allotropes}}, J. Am. Chem. Soc. {\textbf{136}}, 9826 (2014).

\bibitem{r15}
H. J. Xiang, B. Huang, E. Kan, S.-H. Wei, and X. G. Gong, {\it{Towards Direct-Gap Silicon Phases by the Inverse Band Structure Design Approach}}, Phys. Rev. Lett. {\textbf{110}}, 118702 (2013).

\bibitem{r16}
R. Seiwatz, {\it{Possible Structures for Clean, Annealed Surfaces of Germanium and Silicon}}, Surf. Sci. {\textbf{2}}, 473 (1964).

\bibitem{r17}
A. Franceschetti and A. Zunger, {\it{The Inverse Band-Structure Problem of Finding an Atomic Configuration with Given Electronic Properties}}, Nature {\textbf{402}}, 60 (1999).

\bibitem{r18}
J. Lee, I.-H. Lee, and J. Lee, {\it{Unbiased Global Optimization of Lennard-Jones Clusters for N$\leq$201 Using the Conformational Space Annealing Method}}, Phys. Rev. Lett. {\textbf{91}}, 080201 (2003).

\bibitem{r19}
J. Lee, J. Pillardy, C. Czaplewski, Y. Arnautova, D. R. Ripoll, A. Liwo, K. D. Gibson, R. J. Wawak, and H. A. Scheraga, {\it{Efficient Parallel Algorithms in Global Optimization of Potential Energy Functions for Peptides, Proteins, and Crystals}}, Comput. Phys. Commun. {\textbf{128}}, 399 (2000).

\bibitem{r20}
J. Lee, H. A. Scheraga, and S. Rackovsky, {\it{New Optimization Method for Conformational Energy Calculations on Polypeptides: Conformational Space Annealing}}, J. Comput. Chem. {\textbf{18}}, 1222 (1997).

\bibitem{r21}
K. Joo, J. Lee, S. Lee, J.-H. Seo, S. J. Lee, and J. Lee, {\it{High Accuracy Template Based Modeling by Global Optimization}}, Proteins {\textbf{69}}, 83 (2007).

\bibitem{r22}
J. P. Perdew, K. Burke, and M. Ernzerhof, {\it{Generalized Gradient Approximation Made Simple}}, Phys. Rev. Lett. {\textbf{77}}, 3865 (1996).

\bibitem{r23}
G. Kresse and D. Joubert, {\it{From Ultrasoft Pseudopotentials to the Projector Augmented-Wave Method}}, Phys. Rev. B {\textbf{59}}, 1758 (1999).

\bibitem{r24}
G. Kresse, J. Furthm\"{u}ller, {\it{Efficiency of Ab-Initio Total Energy Calculations for Metals and Semiconductors Using a Plane-Wave Basis Set}}, Comput. Mater. Sci. {\textbf{6}}, 15 (1996).

\bibitem{r25}
M. S. Hybertsen and S. G. Louie, {\it{Electron Correlation in Semiconductors and Insulators: Band Gaps and Quasiparticle Energies}}, Phys. Rev. B {\textbf{34}}, 5390 (1986).

\bibitem{r26}
L. Hedin, {\it{On Correlation Effects in Electron Spectroscopies and the GW Approximation}}, J. Phys. Condens. Matter {\textbf{11}}, R489 (1999).

\bibitem{r27}
E. E. Salpeter and H. A. Bethe, {\it{A Relativistic Equation for Bound-State Problems}}, Phys. Rev. {\textbf{84}}, 1232 (1951).

\bibitem{r28}
D. Y. Kim, S. Stefanoski, O. O. Kurakevych, and T. A. Strobel, {\it{Synthesis of an Open-Framework Allotrope of Silicon}}, Nature Mater. {\textbf{14}}, 169 (2014).

\bibitem{r29}
W. Shockley and H. J. Queisser, {\it{Detailed Balance Limit of Efficiency of p-n Junction Solar Cells}}, J. Appl. Phys. {\textbf{32}}, 510 (1961).

\bibitem{r30}
M. A. Green, A. Ho-Baillie, and H. J. Snaith, {\it{The Emergence of Perovskite Solar Cells}}, Nature Photon. {\textbf{8}}, 506 (2014).

\bibitem{r31}
C. Guill\'{e}n, {\it{CuInSe$_2$ Thin Films Grown Sequentially from Binary Sulfides as Compared to Layers Evaporated Directly from the Elements}}, Semicond. Sci. Technol. {\textbf{21}}, 709 (2006).

\bibitem{r32}
M. L. Cohen and B. D. Malone, {\it{Wave Function Engineering: Other Phases of Si for Photovoltaic Applications}}, J. Appl. Phys. {\textbf{109}}, 102402 (2011).

\bibitem{r33}
M. A. Green and M. J. Keevers, {\it{Optical Properties of Intrinsic Silicon at 300 K}}, Prog. Photovoltaics Res. Appl. {\textbf{3}}, 189 (1995).

\bibitem{r34}
L. Yu, and A. Zunger, {\it{Identification of Potential Photovoltaic Absorbers Based on First-Principles Spectroscopic Screening of Materials}}, Phys. Rev. Lett. {\textbf{108}}, 068701 (2012).

\bibitem{r35}
K. C. Pandey, {\it{Reconstruction of Semiconductor Surfaces: Buckling, Ionicity, and $\pi$-Bonded Chains}}, Phys. Rev. Lett. {\textbf{49}}, 223 (1982).

\bibitem{r36}
C. Battaglia, P. Aebi, and S. C. Erwin, {\it{Stability and structure of atomic chains on Si(111)}}, Phys. Rev. B {\textbf{78}}, 075409 (2008).

\bibitem{r37}
J. D. Bryan and G. D. Stucky, {\it{Eu$_4$Ga$_8$Ge$_16$: A New Four-Coordinate Clathrate Network}}, Chem. Mater. {\textbf{13}}, 253 (2001).

\bibitem{r38}
O. O. Kurakevych, T. A. Strobel, D. Y. Kim, T. Muramatsu, and V. V. Struzhkin, {\it{Na-Si Clathrates Are High-Pressure Phases: A Melt-Based Route to Control Stoichiometry and Properties}}, Cryst. Growth Des. {\textbf{13}}, 303 (2013).

\end{references}
\end{document}